\documentclass{pasj00}
\draft

\begin{document}
\SetRunningHead{Author(s) in page-head}{Running Head}

\title{An Investigation into Surface Temperature Distributions of
High-B Pulsars}

\author{Nobutoshi \textsc{Yasutake} } %
\affil{Department of Physics, Chiba Institute of Technology, \\ 2-1-1 Shibazono, Narashino, Chiba 275-0023, Japan}
\email{nobutoshi.yasutake@it-chiba.ac.jp}

\author{Kei \textsc{Kotake}}
\affil{Department of Applied Physics, Faculty of Science, Fukuoka University, \\ 8-19-1 Nanakuma, Jonan-ku, Fukuoka, 814-0180, Japan}\email{kkotake@fukuoka-u.ac.jp\\ Division of Theoretical Astronomy, National Astronomical Observatory of Japan, 2-21-1 Osawa, Mitaka, 181-8588, Japan} 

\author{Masamichi \textsc{Kutsuna}}

\affil{Research Center for the Early Universe (RESCEU), School of Science, The University of  Tokyo, \\
Bunkyo-ku, Tokyo 113-0033, Japan}\email{kutsuna@resceu.s.u-tokyo.ac.jp}

\and

\author{Toshikazu {\sc Shigeyama}}
\affil{Research Center for the Early Universe (RESCEU), School of Science, The University of  Tokyo, \\
Bunkyo-ku, Tokyo 113-0033, Japan}\email{shigeyama@resceu.s.u-tokyo.ac.jp}


%
\def\bmath#1{\mbox{\boldmath $#1$}}


\maketitle

\begin{abstract}
Bearing in mind the application to high-magnetic-field (high-B) 
radio pulsars,
we investigate two-dimensional (2D) thermal evolutions of neutron stars 
(NSs). We pay particular attention to the influence 
 of different equilibrium configurations on the 
 surface temperature distributions. The equilibrium configurations are
 constructed in a systematic manner,
 in which both toroidal and poloidal magnetic fields are determined 
self-consistently with the inclusion of general relativistic effects. 
To solve the 2D heat transfer inside the NS interior out to the crust, 
we have developed an implicit code based on a finite-difference scheme 
that deals with anisotropic
 thermal conductivity and relevant cooling processes in the 
 context of a standard cooling scenario. In agreement with previous 
 studies,
 the surface temperatures near the pole become higher than those 
 in the vicinity of the equator as a result of anisotropic heat 
 transfer. Our results show that the ratio of the highest to
 the lowest surface temperatures changes maximally by one order of magnitude, 
depending on the equilibrium configurations. 
Despite such difference, we find that the area of 
 such hot and cold spots is so small that the simulated X-ray 
spectrum could be well
reproduced by a single temperature blackbody fitting.
\end{abstract}

\section{Introduction}

Over the past few decades, our understanding of 
neutron stars (NSs) has been significantly progressed thanks to
discoveries of several new classes of objects 
(see \cite{kaspi10} for a review). In addition to the conventional
 ``rotation-powered pulsars" (RPPs), great advances in X-ray 
 observations such as by Chandra, XMM Newton, and Swift have led 
 to the discovery of a garden variety of isolated NSs, which include
 magnetars (e.g., \cite{woods04,mere08} for reviews), high-magnetic-field (high-B) 
pulsars (e.g., \cite{ng11,ng12}), X-ray-isolated neutron stars (XINSs,
 see \cite{haberl07} for a review), 
and central compact objects (CCOs, e.g., \cite{gotthelf13}).  Among them, an extreme class is magnetars including
 Anomalous X-ray Pulsars (AXPs) and Soft $\gamma-$ray Repeaters (SGRs),
 which have very large estimated magnetic fields ($B \sim 10^{14} 
- 10^{15}$ G) and exhibit violent flaring activities (see \cite{rea11}
 for a review). Such high magnetic fields are 
believed to be responsible for explaining the observational 
characteristics (e.g., \cite{thomp95,thomp01}), however
 the origin of magnetars (whose strong fields either come 
 from the postcollapse rapidly spinning NSs \citep{thompson93}
or descend from the main sequence stars \citep{ferrario}) and 
the relation to the more conventional RPPs have not yet been 
 clarified.

The big gap between these two classes of objects has been bridged 
thanks to the recent discoveries of a weakly magnetized magnetar
 (SGR 0418+5729; \cite{rea10,turo11}), magnetar-like bursts from a 
rotation-powered pulsar (PSR J1846|0258 \citet{gavriil08,ng08}), and 
pulsational
radio emission from magnetars \citep{camilo06,camilo07,rea12}. These pieces of observational evidence  lends support to a unified vision of 
NSs \citep{kaspi10} that magnetars and the conventional RPPs could 
originate from the same population
(see \citet{ng12,olausen13} for collective references therein).
 In this context, 
high-B pulsars are attracting a paramount attention, 
which is very likely to connect the X-ray quiet standard RPPs with very active magnetars, showing intermediate luminosities, and  occasional magnetar-like activities \citep{perna11}.
 
In order to understand what is the underlying physics leading to the unification theory of NSs, it is of 
primary importance to calculate the structure and evolution of the NSs, and compare a theoretical model with observational data. Extensive studies have been performed so far
 in a variety of contexts (e.g., \cite{yakovlev04,page06} for reviews).
One important lesson we have learned from accumulating observations \citep{zavlin07, haberl07, nakagawa09} is that the surface temperature of isolated NSs is not spherically symmetric (1D). This demands us to go beyond 1D modeling (e.g., \cite{greenstein,nomoto86,dany92,potekhin01}, and 
see \cite{chris92} for collective references therein) to multi-dimensional (multi-D) modeling for the evolutionary calculations.

As has been understood since \citet{greenstein}, the presence 
 of a sufficiently strong magnetic field ($B \gtrsim 10^{10}$ G),
 ubiquitous such as in the envelope of a NS, leads to anisotropy of heat transport due to both classical and quantum magnetic field effects. 
As a result, electron thermal conductivity is strongly suppressed 
 in the direction perpendicular to the magnetic field and increased along 
 the magnetic field lines \citep{canuto70,itoh75}, which makes the regions around the magnetic poles warmer than those around the magnetic equator (the so-called {\it heat blanketing} effect). To accurately understand the origin of the observed surface temperature anisotropy, multi-D 
(currently limited to axisymmetric two-dimensional (2D)) 
calculations of a NS have been performed extensively so far (e.g., \cite{geppert02,geppert04,perez06,pons07,aguilera08,
pons09,vigano13}, and see collective references therein). 

One of numerical difficulties of the multi-D evolutionary models comes
from the fact that one needs to deal with 
various dissipation processes of magnetic
 fields working over several orders of magnitudes in the
 physical scales during the long-term evolution. \citet{goldreich92} 
were the first to identify the dissipation processes 
of the magnetic energy in the crust of an isolated NS during its 
evolution.
 On top of the ohmic decay and the ambipolar diffusion, they first proposed that the Hall drift, though non-dissipative itself, could be an important ingredient for the field decay because it can lead to 
dissipation through a whistler cascade of the turbulence. 
To unambiguously understand how the dissipation proceeds, multi-D 
numerical simulations focusing on electron MHD equations (EMHD, e.g., \citet{Biskmap,Cho04,Cho09,takahashi11}) are required, because the turbulent cascade from 
 large to small scale inherent to the Hall term is essentially a non-linear process. 

In step with these advances in microphysical simulations shedding 
 light on the physics of
 the dissipation processes, 2D magneto-thermal evolutionary simulations
 have been developed with increasing sophistication, 
in which best neutrino processes (e.g., \cite{yakovlev01})
 and thermal conductivities currently available are implemented (e.g., \citet{vigano12} for a review). 
Most recent code based on finite-difference schemes
 \citep{vigano12} can handle arbitrarily large magnetic 
fields with the inclusion of the Hall term, which had been a 
big challenge in the previous code employing a spectral method 
(e.g., \citet{hollerbach02}). By computing an extensive set of such state-of-the-art evolutionary models, \citet{vigano13} recently 
pointed out that the mentioned impressive diversity of 
 data from X-ray space missions can be explained by variations of NS's initial magnetic field, mass and envelope composition, which is well consistent with the concept
 of the unification scenario.

Joining in these efforts, we investigate 2D thermal 
evolution of NSs in this study. Having in mind the application 
 to high-B pulsars, we pay attention to the influence of 
 different equilibrium configurations of NSs on the surface 
 temperatures distributions. Since the Hall term plays an important 
 role in the evolution of very 
large magnetic field ($\gtrsim 10^{14}$ G, e.g., \citet{vigano13}),
 we only take into account 
 the effects of magnetic field decay via a simplified analytical 
 prescription (e.g., \citet{aguilera08}).
 To solve 2D heat transport inside 
 the NS interior out to the crust, we develop an implicit 
code based on a finite-difference scheme, which deals with anisotropic 
thermal conductivity and relevant 
 microphysical processes in the context of a standard cooling 
scenario \citep{yakovlev01}. The equilibrium configurations of 
 NSs are constructed in a systematic manner by employing 
 the Tomimura-Eriguchi scheme \citep{tomimura05}, in which both toroidal
 and poloidal magnetic fields can be determined self-consistently
 with the inclusion of general relativistic effect \citep{kiuchi08a}.
 We employ a nuclear equation of state based on the relativistic
 mean-field theory by \citet{shen98}.
 Based on the surface temperature distributions, we discuss
 the properties of X-ray spectrum expected from a variety of our 
2D models\footnote{For simplicity, we do not consider non-thermal X-ray emission such as in the case of PSR J1846-0258~(\cite{livingstone11})}.


This paper is organized as follows. In Section~II, we outline our initial models and numerical methods for thermal evolutions. In Sec. III, we present numerical results. Section IV is devoted to the summary and discussion.
 
\newpage

\section{Numerical Methods}

\subsection{Equilibrium configurations}
\label{subsec:2-1}

\begin{figure}[tbh]
\begin{center}
\includegraphics[width=80mm]{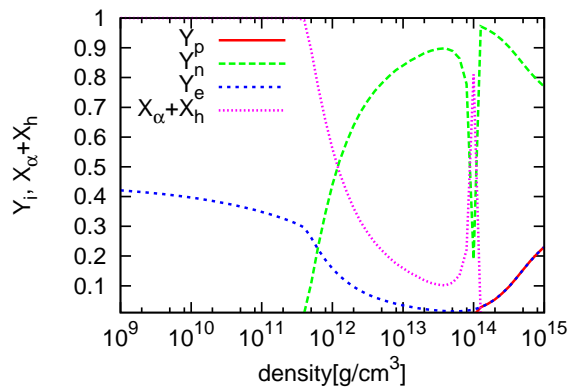}
\caption{
(Color online) Compositions for Shen EOS~(\cite{shen98}); i.e. proton fraction $Y_p$ (solid line), neutron fraction $Y_n$ (dashed line), electron fraction $Y_e$ (short-dashed line). The sum of fraction of alpha particle $X_\alpha$ and  heavy nuclei  of $X_h$ is indicated by dotted lines. }
\label{fig:01}
\end{center}
\end{figure}

The magnetic field configuration in the interior of NSs is 
poorly known from observations. Hence we adopt stationary states
 of magnetized stars as the initial conditions for our 2D evolutionary
 calculations. We employ a nuclear EOS by \citet{shen98} that 
 is based on the relativistic mean-field theory. Fig. \ref{fig:01} 
shows the number fraction of each element as a function of density.
 Note here that zero temperature is assumed for the case of 
cold NSs ($T = 0$).  As we will discuss later, 
the composition and density distribution are the 
critical factors to determine the cooling processes.
 For simplicity, we leave the inclusion of 
 hyperons as well as pions, kaons, quarks as our future work, although
 some recent studies suggest 
 the existence of hyperon matter 
(\cite{weissenborn12a, weissenborn12b}) can explain $\sim 2 M_{\odot}$ NSs (\cite{demorest10,antoniadis13}).


Employing the Shen EOS, we construct equilibrium stellar 
configurations. The basic equations and the numerical 
methods for this purpose are already given in
(\cite{tomimura05,yoshida06,yoshida06b}, see also \cite{kiuchi08a}). 
Hence, we only give a brief summary for later convenience.

Assumptions to obtain the equilibrium models are summarized as follows.
(1) Equilibrium models are stationary and axisymmetric. 
(2) The matter source is approximated by a perfect fluid with infinite conductivity. Note that this assumption is valid before magnetic field begins to decay on a timescale of $10^{6}$ yr \citep{pons07prl}. 
 (3) There is no meridional flow of matter. 
 (4) The magnetic axis and rotation axis are aligned. 


With these assumptions, we need to specify
 some arbitrary parameters in the above scheme for determining
 equilibrium configurations.
First, the toroidal magnetic field has a functional form with respect to
 the so-called flux function $u$ as
\begin{eqnarray}
\label{eq:01}
B_{\rm \phi}= \frac{a (u-u_{\rm max})^{k+1}  } {R^2 (k+1)},
\end{eqnarray}
where $a$ and $k$ are arbitrary constants to determine the magnetic field strength, and $u_{\rm max}$ is the maximum value of the 
flux function $u$ that depends on $R$ and the vector potential $A_\phi$ 
 as $u= R A_\phi$. Note we employ the cylindrical coordinates $(R, \phi,
 z)$. Another important parameter appears in the equation of 
 current density $j^a$,
\begin{eqnarray}
\frac{j^a}{c} = \frac{\kappa}{4 \pi} {B^a} +  \rho 
(\mu + R^2 \Omega \Omega') \varphi^a,
\end{eqnarray}
 where $\mu$ is the input parameter, and other variables
 ($\kappa$, $\Omega\,$; angular velocity, $\rho\,$; mass density) can 
 be determined\footnote{$\varphi^a$ represents the rotational Killing vector.} self-consistently by solving the generalized Grad-Shafranov equation (equivalently the Maxwell equations for
 constructing the equilibrium configurations) and time-independent 
Euler equations (see \cite{tomimura05, yoshida06, kiuchi08a} for more details). The only remaining parameter is the central density of a star 
($\rho_c$).

 In Table 1 we summarize the four parameters ($a$, $\mu$, $k$ and $\rho_c$)
 to obtain the equilibrium configurations. In the table,  
models M000 and m000 are non-magnetized models ($a = 0$). The 
 difference between them is the central density, which is higher 
for model M000 leading to greater (baryonic) mass 
($2.20 M_{\odot}$, see Table 2) than for model m00 ($1.67 M_{\odot}$).
As is well known, if a maximum density of a star is higher than
 the nuclear saturation density $\rho_{\rm nuc} \approx 2.8 \times
 10^{14}~{\rm g}~{\rm cm}^{-3}$, one needs to take into account 
a general relativistic (GR) effect. However, the fully GR
 approach to the magnetized equilibrium configuration has 
 not been established yet except for the purely poloidal \citep{bocquet95,cardall01}
 or toroidal fields
\citep{kiuchi08b}. Therefore we employ an approximate,
 post-Newtonian method proposed by \citet{kiuchi08a}.

 Back to Table 1, models mauk, maUk, MaUk and mAUk, are all 
magnetized models. In Table~\ref{tab:02}, several important quantities of 
the equilibrium models are summarized, i.e., 
the mass $M$, the radius $R$, the poloidal magnetic field at the 
pole $B_p$, the magnetic filed in the center $B_c$, the ratio of the 
total magnetic field energy to the gravitational energy $H/|W|$, and 
the ratio of the maximum magnetic field for the toroidal component to
 that for the poloidal one $B_{\phi, \rm{max}}/B_{p, \rm{max}}$. 
For all the magnetized models, $B_p$ is set to take $\sim 10^{13}$ G, 
 which is reconciled with high-B pulsars. Since $H/|W|$ are quite small
($\ll 1$), the equilibrium configurations are not affected by the 
 Lorentz force (see, however \citet{yasutake10} for ultra-strong field 
($B \gtrsim 10^{17}$ G) case). The ratio of 
$B_{t, \rm{max}}/B_{p, \rm{max}}$ is in line with results from recent 
MHD simulations (\cite{braithwaite04}), showing
 that stable configurations require 
 the coexistence of both poloidal and toroidal components, 
 approximately of the same strength.
As an experimental point of view, we compute two extreme cases for models maUS and mSUK, which do not satisfy the stability condition.

We checked the convergence of the presented results by doubling
 the number of mesh points from the standard set of radial and angular
 direction mesh points of 100 $\times$ 100. 
We set the uniform zones in the polar direction while non-uniform 
zones in the radial direction to describe the density profile and 
the particle compositions in the crusts precisely. Here, the 
zone-interval in the radial direction is, $dr_n = dr_0 \times q^{n-1}$ from the center, where the indent $n$ is the zone number measured from the 
center. The constant $dr_0$ denotes the maximum grid interval that is 
 related to the maximum equatorial radius $R_{\rm eq}$ of each model as 
$dr_0 = ( R_{\rm eq} - 1 ) / (n_{\rm max} -1)$. 
Here, $n_{\rm max}$ denotes the maximum zone-number, set as $n_{\rm max} = 100$ and $q=0.99$ in this study. The minimum grid interval is of the 
order of one meter in all our simulations, which is small enough to calculate the crust of a NS. By checking 
 the virial identities \citep{cowling65} for all the models, we confirm 
 that the typical values are of the orders of magnitude
 $\lesssim 10^{-3}$, which are almost the same for the polytropic
 EOS case \citep{tomimura05,yoshida06}. In general, 
the convergence is known to become much worse for realistic EOS because
 their adiabatic index is not smooth especially near $\rho_{\rm nuc}$.
In this respect, our numerical scheme works well.

\begin{table}
\caption{\label{tab:01}
Models with parameters to determine magnetic field distributions. Note that the unit of $a$ is depends on the value of $k$ through Eq.(\ref{eq:01}). As for the values of $a$, $\mu$, $k$ in this table, we adopt the geometrical units as in previous studies \citep{tomimura05, yoshida06, kiuchi08a}. 
}
\begin{center}
\begin{tabular}{lccccc}
Models &  $\rho_c$    & $a$ & $\mu$  & $k$ &\\
            & [10$^{14}$ g cm$^{-3}$] &          &                &                     &\\
\hline   
   M000    & 12.00 &   0.0    &  0.00 $\times 10^{0}$ & 0.00  &  \\
   m000    &   6.00 &   0.0    &  0.00 $\times 10^{0}$ & 0.00  &  \\
\hline
   mauk     &    6.00 & 70.0   & 1.00 $\times 10^{-5}$ & 1.00$\times 10^{-1}$  &  \\
   maUk    &    6.00 & 70.0    & 1.00 $\times 10^{-4}$ & 1.00$\times 10^{-1}$  &  \\
   MaUk     & 12.00 & 70.0    & 1.00 $\times 10^{-4}$ & 1.00$\times 10^{-1}$  &  \\
   mAUk    &    6.00 & 100.0 & 1.00 $\times 10^{-4}$ & 1.00$\times 10^{-1}$  &  \\
\hline   
   maUS  &    6.00 & 70.0    & 1.00 $\times 10^{-4}$ & 1.00$\times 10^{0}$   &  \\  
   mSUK   & 6.00 & 5000.0 & 1.00 $\times 10^{-4}$ & 8.00$\times 10^{-1}$  &  \\  
\end{tabular}
\end{center}
\end{table}

\begin{table*}[hbt]
\caption{\label{tab:02}
Summary of initial Models (see text for the definition of variables).}
\begin{center}
\begin{tabular}{lcccccc}
Models & $M$ & $R$ & $B_p$ & $B_c$  & $H/|W|$ & $B_{t, \rm{max}}/B_{p, \rm{max}}$ \\
            & [$M_\odot$] & [km] &[10$^{14}$ G]    &  [10$^{14}$ G ] & & \\
\hline
    M000    & 2.20 & 12.9 & - & - & - & - \\
   m000    & 1.67 & 14.3 & - & - & - & - \\
 \hline
   mauk     & 1.67 & 14.4 & 0.05 & 0.21 & 1.09 $\times 10^{-10}$ & 0.57 \\
   maUk   & 1.67 & 14.3   & 0.48 & 2.11 & 1.26 $\times 10^{-8}$ & 0.75 \\
   MaUk     & 2.20 & 12.9 & 0.72 & 3.22 & 8.49 $\times 10^{-9}$ & 0.65 \\
   mAUk    & 1.67 & 14.2 & 0.45 & 2.26  & 1.49 $\times 10^{-8}$ &1.04 \\
 \hline
   maUS   & 1.67 & 14.2 & 0.42 & 1.76& 5.73 $\times 10^{-9}$ & 2.80$\times 10^{-6}$\\  
   mSUK & 1.67 & 14.3 & 0.44 & 1.84 & 6.30 $\times 10^{-9}$ & 0.13 \\  
 \end{tabular}
\end{center}
\end{table*}

\begin{figure}[htb]
\begin{center}
\includegraphics[width=80mm]{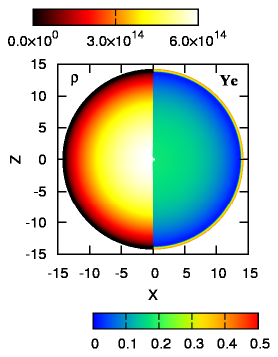}
\caption{(Color online) The density profile $\rho$[g cm$^{-3}$], and the electron fraction $Y_e$  for model maUk in Table~I. Note that the distributions are all similar to the remaining models m000, mauk, mAUk, and maUS.}
\label{fig:02}
\end{center}
\end{figure}

Figure \ref{fig:02} shows equilibrium configuration for our fiducial model (maUk). As already mentioned, the assumed field strength is not so dynamically strong that the density and composition distributions are essentially spherical ($M$ and $R$ in Table 2 are hardly dependent on the field strength). Therefore the density distribution for model maUK is similar to the remaining models, given the same central density\footnote{For all the models, the minor to major axis ratio is set as 0.99}.

\begin{figure*}[htb]
\begin{center}
\includegraphics[width=80mm]{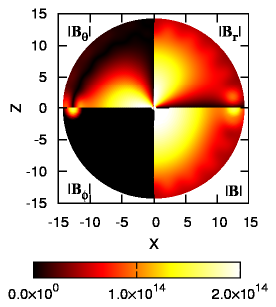}
\includegraphics[width=80mm]{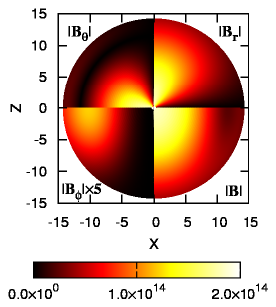}
\caption{
(Color online) The strengths of initial toroidal, poloidal, and total magnetic field; $|B_\phi|$, $|B_p|$, and $|B|$ in unit [10$^{14}$ G] (see
 text for the definition) for model maUk (left panel) and
 mSUK (right panel), respectively. Note in the right panel 
that the strength of toroidal magnetic field, $|B_\phi|$, 
is multiplied by a factor of 5 for visualization. 
}
\label{fig:03}
\end{center}
\end{figure*}

The left and right panels of Figure~\ref{fig:03} show the distribution 
of magnetic field for model maUK and mSUK, respectively, 
 with respect to the lateral ($|B_{\theta}|$, left top),
 azimuthal ($|B_{\phi}|$, left bottom), radial component ($|B_{r}|$, right top), and the sum ($B = \sqrt{B_{r}^2 + B_{\theta}^2 + B_{\phi}^2}$, 
right bottom, all in the absolute value).
Note again that the magnetic distribution is dependent on 
the four parameters ($\rho_c$, $a$, $\mu$, and $k$). 
From the right panel, it can be seen that the toroidal magnetic 
field ($|B_{\phi}|$ amplified by a factor of 5, bottom left panel) 
is dominant over the poloidal components
(e.g., the top panels for $|B_{\theta}|$ and $|B_r|$)
 in the vicinity of the equatorial plane out to $\sim 7$ km in radius.
The dominance of the toroidal fields in the vicinity of the outer 
 regions is common to model maUk (right panel of Figure 3).
 The most remarkable difference between the left and right panel
 in Figure 3 is the distribution of the 
toroidal component $|B_\phi|$, which is confined in a narrow
 region for model maUk (seen as a half-circle colored by yellow in the 
$|B_{\phi}|$ plot) at a radius of $\sim 14$ km in the vicinity of the equatorial plane. This difference comes from the parameter $k$.
 Smaller $k$ makes the distribution of $|B_\phi|$ more compact
 as seen in the left panel. This feature is common to models 
 with smaller $a$ in Table 1, that is, mauk, MaUk, and maUS.

\subsection{Thermal evolution}
Taking GR effects into account, we employ a spherically symmetric 
 metric given by the equation \citep{misner}
\begin{eqnarray}
ds^2 = -e^{2\phi} dt^2 + e^{2\Lambda} dr^2 + r^2d\Omega^2.
\end{eqnarray}
Under this background metric, the thermal 
 evolution of a NS can be described by the energy 
balance equation (e.g., \cite{aguilera08}),
\begin{eqnarray}
\label{eq:evo}
c_v e^\phi \frac{\partial T}{\partial t} + \nabla \cdot (e^{2\phi} {\bmath F}) = e^{2\phi}(H-L),
\label{eq:evo}
\end{eqnarray}
where $c_v$ is the specific heat capacity,
 $T$ is the temperature, $H$ and $L$ is the energy loss and gain by neutrino emission and by the 
Joule heating, respectively. In the diffusion limit,
 the heat flux (${\bmath F}=(F_r, F_\theta)$) can be expressed as
\begin{eqnarray}
\label{eq:fr}
{F_r} &=& -e^{-\phi} (\kappa_{rr} e^{-\Lambda} \partial_r \tilde T
                                     +\frac{\kappa_{r\theta}}{r} e^{-\Lambda} \partial_\theta \tilde T),\\
\label{eq:fth}
{F_\theta} &=& -e^{-\phi} (\kappa_{\theta r} e^{-\Lambda} \partial_r \tilde T
                                     +\frac{\kappa_{\theta \theta}}{r} e^{-\Lambda} \partial_\theta \tilde T),
\end{eqnarray}
where $\kappa$ is the total thermal conductivity tensor and 
 $\tilde T$ is the red-shifted temperature ($\tilde T = e^\phi T$). 
 The dominant contribution to $\kappa$ comes from electrons (see \citet{geppert04,page07,aguilera08}), which can be written as 
\begin{eqnarray*}
\label{eq:kappa}
\kappa_e& =& \kappa_e^\perp 
\Biggl[
{\bmath E}  \\
&+&  (\omega_B \tau)^2
\left ( \begin{array}{ccc}
b_{rr}             & b_{r \theta}           & b_{r \phi}  \\
b_{\theta r} & b_{\theta \theta} & b_{\theta \phi} \\
b_{\phi r}      & b_{\phi \theta}     & b_{\phi \phi} 
\end{array}  \right )
+  \omega_B \tau
\left ( \begin{array}{ccc}
0                   & b_{\phi}  & - b_{\theta}  \\
-b_{\phi}    & 0               & b_r \\
b_{\theta} & -b_r         & 0
\end{array}  \right )
\Biggr],
\end{eqnarray*}
where $\kappa_e^\perp$, $\omega_B$, and $\tau$ are the electron 
thermal conductivity orthogonal to the magnetic field, the gyro-frequency ($\omega_B = eB/m_e^* c$ with $m_e^*$ being the effective electron mass), 
and the electron relaxation time \citep{urpin80}. 
Here, $\bmath{E}$ is the identity matrix, and $b_r$, $b_\theta$, $b_\phi$ 
denotes the radial, lateral, and azimuthal component
 of the unit vector in the direction of the magnetic field, respectively.
 We employ a public code to calculate these kinematic coefficients
 in the crust\footnote{www.ioffe.rssi.ru/astro/conduct/condmag.html}. 
 For the inner core, we adopt the formula in \citet{gnedin95}.

Concerning the heat capacity ($c_v$ in Eq. (4)), we assume that electrons are degenerate, and baryons are non-relativistic (\cite{aguilera08}).
In the crust, the heat capacity is provided by electrons, ions, and free 
neutrons. We ignore the ion contribution on the heat capacity because the contribution is small. For simplicity, 
we do not take the effect of superfluidity into account.

Regarding the cooling term $L$ in Eq.(\ref{eq:evo}), 
we follow the so-called standard cooling 
 scenario (\cite{yakovlev01} for a review), in which 
the total cooling rate is dominated by slow processes in the core,
 such as by modified Urca and nucleon-nucleon Bremsstrahlung
 \citep{yakovlev95, haensel96}. As a first step, we think the 
consideration of the minimal cooling scenario \citep{page04} or 
enhanced cooling scenario \citep{lattimer91,prakash92,takatsuka04} 
as an important
 extension as a sequel of this study.

As for the magnetic field decay,
 we only take into account the Ohmic dissipation, because
 in our case ($B_0 \sim 10^{13}$ G) the Hall term plays an only 
minor role. For simplicity, we assume 
 that the field geometry is fixed and the evolution is included 
 only in the normalization $B$ as \citep{aguilera08}
\begin{eqnarray}
B = B_0\,{\rm exp}\,(-t/\tau_{{\rm Ohm}}),
\end{eqnarray}
 where $B_0$ is given by the initial equilibrium configurations,
 and the Ohmic decay timescale ($\tau_{{\rm Ohm}}$) is set 
as $10^6$ yr \citep{pons07prl} in all the models.
 Simple as it is, such a prescription is known to be able
 to reproduce qualitatively results from more detailed simulations
 \citep{pons07}.
The heating rate ($H$) in Eq.(4) is given by the integral of 
$H = \int \Delta B^2/8\pi ~dV$, where $\Delta B \equiv B^{n+1} - B^{n}$ 
denotes the decrease of the field strength in each computational 
timestep (between $n+1$ and $n$ steps). Note that for a middle-age NS
 of $10^3 - 10^5$ yr, to which we pay attention in this work, the 
 Ohmic decay does not significantly affect the thermal evolution.

Here, let us estimate the diffusion timescale $\tau_D$ from
 Eq. (\ref{eq:evo}). $\tau_D$ is proportional to
 $\sim c_v' (\Delta x)^2 /\kappa' $, where $c_v'$, $\Delta x$, and $\kappa'$ are typical values for the heat capacity, the grid interval, and the thermal conductivity. Taking typical values in the core of a NS, 
such as $c_v' \sim 10^{20}$ erg cm\,$^{-3}$\,K$^{-1}$, $\Delta x \sim 10^3$ cm, and $\kappa' \sim 10^{23}$ erg K$^{-1}\,{\rm s}^{-1}\,{\rm cm}^{-1}$,
the diffusion time scale is estimated as $\tau_D \sim 10^3$ s.
Since the evolution timescale of a NS ($\sim 10^4 - 10^6$ years) 
 is much longer than the diffusion timescale, an implicit scheme is 
 needed to solve Eq. (4). In doing so, we take an operator 
splitting method. The second term in Eq. (\ref{eq:evo}) 
includes the cross terms of second derivatives such as
 $\partial_r \partial_\theta$ and $\partial_\theta \partial_r$, 
which is not straightforwardly handled by a standard 
matrix inversion scheme. We treat these terms 
as a source term to get a convergent solution (see Appendix A 
for more details). Finally, we employ a phenomenological
 formula to estimate the surface temperature $T_s$ from the temperature
 at the bottom of the envelope $T_b$ \citep{potekhin01} (e.g., Eq. (31,32) in \citet{aguilera08}), in which the density at the 
 bottom of the envelope is set as $10^9$ g cm$^{-3}$.

\section{Result}

\begin{figure*}[hbt]
\begin{center}
\includegraphics[width=80mm]{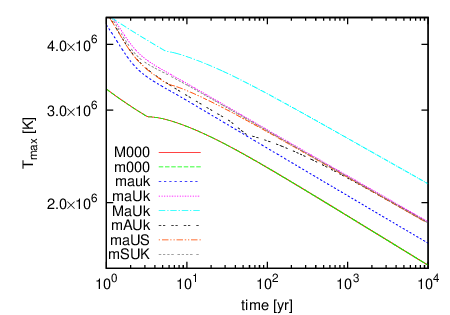}
\includegraphics[width=80mm]{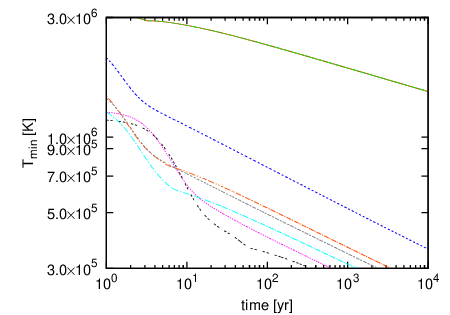}
\caption{(Color online) 
Cooling light curves for all the computed models. Left and right 
panel shows the evolutions of the maximum ($T_{\rm max}$) and minimum
($T_{\rm min}$) surface temperatures (see text for more details).
Note the difference of the temperature scale in each panel.
}
\label{fig:evo}
\end{center}
\end{figure*}

\begin{figure*}
\begin{center}
\includegraphics[width=80mm]{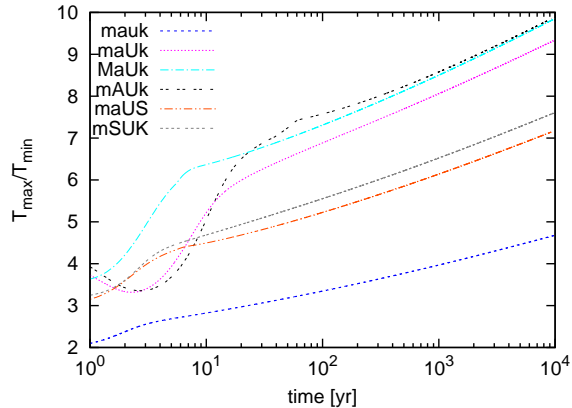}
\caption{Time evolution of the ratio of $T_{\rm max}$ to $T_{\rm min}$ 
for all the magnetized models.}
\label{fig:ratio}
\end{center}
\end{figure*}


Due to the mentioned heat blanketing effect, the surface temperature
 has its maximum ($T_{\rm max}$) and minimum ($T_{\rm min}$) 
in the vicinity of the magnetic poles and equator for our 2D evolutionary models with magnetic fields. The left  and right panels of Figure \ref{fig:evo} show the evolution of 
$T_{\rm max}$ and $T_{\rm min}$ for all the computed models, respectively.
 Since the characteristic age for high-B pulsars is estimated 
as $10^{3-5}$ yr from observations (e.g., \citet{ng12}), we focus on
 the evolution up to $t_{\rm end} = 10^{4}$ yr in this paper.
 Before going into detail, let us compare our results with 
 \citet{aguilera08}, especially with their Figure 13 (central panel)
 for their PC model, in which a similar field strength to our model maUk 
 is employed with the use of the standard cooling processes.
The temperature of the hot spot for our model maUk (e.g., 
pink dashed-line in the left panel of Figure 4) drops
 about 50 \% from the birth ($T_{\rm max}$ = $4.4 \times 10^{6}$ K) 
to the age of $10^3$ yr ($T_{\rm max} = 2.4\times 10^{6}$ K), while 
the PC model \citet{aguilera08} drops about 40 \% for the same timescale
(e.g., from $T_{\rm max} \sim 10^{6.5}$ K to $\sim 10^{6.1}$ K, purple line
 in their figure). Therfore, our results are in good agreement with 
 previous study, granted that similar field strength and microphysics are employed (e.g., \citet{aguilera08}).

Figure 5 shows that the contrast ratio of $T_{\rm max}$ to $T_{\rm min}$ 
for the magnetized models at $t_{\rm end}$ ranges from $\sim$ 5-10, 
while the contrast ratio is unity for non-magnetized 
 models (M000, m000). 
In our magnetized models, the lowest and highest contrast ratio (4.86 
and 10.0) is obtained for our most weakly and strongly
 magnetized model (mauk and mAUk), respectively (e.g., 
$H/|W|$ in Table \ref{tab:02}). Note that these values are 
quantitatively in good agreement with previous results 
in which more detailed 
 cooling processes and heat transport scheme than ours were
 employed (\cite{aguilera08}). It should be mentioned that the contrast ratio does not 
depend solely on the magnetized parameter $H/|W|$ but also on the configuration of the magnetic field. 
For example, the contrast ratio for model MaUk is 
larger than that for model maUk, although $H/|W|$ is 
smaller for model MaUk. 

Note here that 
the initial mass hardly affects the contrast ratio. 
In fact, the surface 
temperatures for models m000 and M000 are degenerate
 as seen from the both panels in Fig \ref{fig:evo}. 
 This is not surprising because we assume the standard 
cooling scenario. The implementation of another cooling scenario may break the degeneracy, but this is 
 beyond the scope of this work as we mentioned 
 earlier. Note also that even
 without the magnetic fields, rapid rotation can lead 
to anisotropic surface temperature distributions
\citep{negreiros12}, which also needs further investigation.


 In the above results, we paid attention only to the surface
 temperatures in the hot ($T_{\rm max}$) and cold ($T_{\rm min}$) spots,
 but the average temperature should be between them. 
 Since the surface temperature is often estimated by 
 a black-body fitting to the observed X-ray spectrum, we move 
 on to calculate spectra from our 2D evolutionary models.
Recently, some observations suggest that 
a two temperature black-body (2BB-) fitting (i.e., cold $\bar{T}_C$ 
and hot $\bar{T}_H$ component) is needed
 to explain observed spectra for magnetars \citep{nakagawa09}. 
 Based on our results, we exploratory discuss how the spectrum could be 
 in the case of high-B pulsars.

\begin{figure}[hbt]
\begin{center}
\includegraphics[width=80mm]{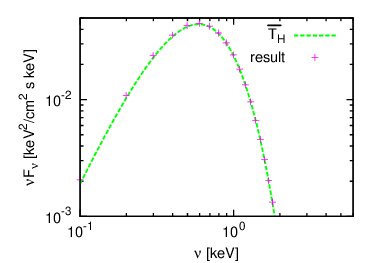}
\includegraphics[width=80mm]{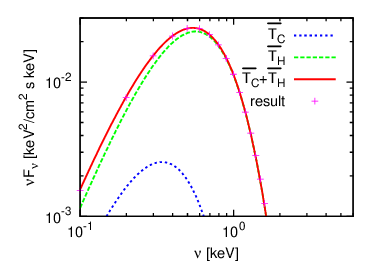}
\caption{(Color online) 
Expected X-ray spectrum calculated for models maUk (left panel)
 and mSUK (right panel) after 
10$^4$ yr. Here we assume the inclination angle from the 
equatorial plane of observer is $\theta = 0^\circ$. 
The dashed lines labeled as ``result" show the estimated spectra from 
 our 2D evolutionary models (see text for more details). The lines 
labeled as $\bar{T}_C$, $\bar{T}_H$, $\bar{T}_C+\bar{T}_H$), 
are the spectra of the hot spot, the cold spot, and the sum
 estimated by a single- or two- temperature BB fitting to
 the cross dotted points (see the text for more details). }
\label{fig:08}
\end{center}
\end{figure}

By integrating the surface fluxes with an assumption of the Plank law, Fig. \ref{fig:08} shows
 the simulated X-ray spectrum for model maUk (left panel) 
and mSUK (right panel) at 10$^4$ yr. We assume the 
 distance to the source as $D$= 1.0 kpc in the following. 
In the figure, we set the inclination angle $\theta$ as 
$\theta=0^\circ$, namely an equatorial observer is assumed.
We will discuss the dependence of the inclination angle later.

In the figure, the cross dotted points denote the results calculated 
from our 2D evolutionary models (labeled as ``result") that include 
 contribution to the spectrum from all the regions on the NS surface.
 To reconstruct the total spectrum (cross dotted points) either from a single or two temperature BB fitting,
 the dashed green line ($\bar{T}_H$), the dashed blue line ($\bar{T}_C$), and the sum ($\bar{T}_H + \bar{T}_C$) represents the spectrum for each component. 
 Here, we determine $\bar{T}_C$ and $\bar{T}_H$ to get the best 
 $\chi$-square fitting to the total spectrum (labeled by ``result")
by changing the area-weighted spectrum with respect to the cold
 (blue dashed line) and hot (green dashed line) component.
  The left panel shows that the single BB-fitting ($\bar{T}_H$) is 
enough to reconstruct the X-ray spectrum for model maUK, although 
  $T_{\rm max}$ is almost 10 times higher than $T_{\rm min}$ as shown 
in Fig \ref{fig:ratio}.  
These features are also similar to other models (mauk, MaUk and mAUk).
 This indicates that the X-ray spectrum from our standard 2D models
 (with the coexistence of both poloidal and toroidal components)
 and also with the assumption of the standard cooling processes can be 
reconstructed by a single BB fitting. On the other hand, the right panel is from one of our 
extreme cases (model mSUK), which shows that the two component
 BB fitting is needed. 

\begin{figure*}[hbt]
\begin{center}
\includegraphics[width=80mm]{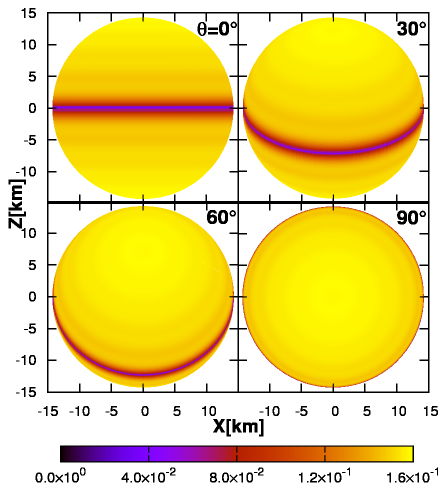}
\includegraphics[width=80mm]{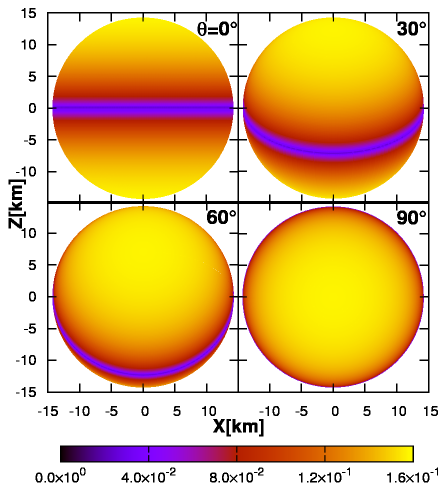}
\caption{(Color online) 
Surface temperature in unit of keV 
for models maUk (left panel) and mSUK (right panel)
 after 10$^4$ years as a function of the viewing angle $\theta$.
}
\label{fig:09}
\end{center}
\end{figure*}

To better understand the reason, we present in Fig. \ref{fig:09}
 the surface temperature distribution for a pair of models 
 (model maUk (left panel ) and mSUK (right panel)). In both of the 
models, cold (colored by blue in the equator) and hot (yellow) 
spots can be seen, however, the area of the cold spot is confined
 in much narrow region for model maUk (left) than for model
 mSUK (right).
  As already mentioned, 
 this is because the toroidal magnetic field $B_\phi$ 
is confined in a narrow region for model maUK (e.g., in the left panel 
Fig.~\ref{fig:03}), which is vice versa for model mSUK (the right panel).
 This is the reason why the spectrum from model maUk can be 
 well fitted by a single temperature component, while two component
 fitting is needed for model mSUK.
 Our results present supporting evidence for previous investigations
(\cite{perez06,geppert06})
that the temperature distribution depends primarily on the magnetic field 
distribution. 

 Table \ref{tab:S_ratio} summarizes the ratio of the surface area 
of the hot regions to the total surface area 
as a function of the inclination angle ($\theta=0^\circ, 30^\circ, 60^\circ,$ and $90^\circ$) for all the models. 
Clearly, our four standard models (mauK, maUk, MaUk, mAUk) show
 a clear dominance of the hot areas over the cold areas (the 
ratio being greater than 0.5), while 
 the ratio approaches to 0.5 for the two extreme cases (models
 maUS and mSUK) especially seen from equator ($\theta = 0^{\circ}$). 
This again indicates that the simulated spectrum from 
our 2D models can be represented by a single temperature BB fitting.
 On the other hand, if the two temperature BB fitting would be required
 for a high-B-pulsar class field strength 
($B_{\rm surface} \approx 10^{13}$ G), it might suggest the 
 existence of larger cold spots (models maUS and mSUK). We speculate that 
this could 
 possibly give some hints to the intrinsic field configuration 
(e.g., Figs.
3 and 7).


\begin{table}[hbt]
\caption{\label{tab:S_ratio}
The ratio of the area of the hot spot to that of the total 
 NS surface.}
\begin{center}
\begin{tabular}{lcccc}
Name & $0^\circ$ & $30^\circ$ & $60^\circ$ & $90^\circ$ \\
\hline   
   mauk &0.807&0.820&0.879&0.879\\
   maUk &0.854&0.866&0.912&0.938\\
   MaUk &0.715&0.722&0.759&0.801\\
   mAUk &0.889&0.901&0.939&0.971\\
\hline   
   maUS &0.581&0.612&0.733&0.782\\  
   mSUK &0.553&0.588&0.717&0.769\\
\end{tabular}
\end{center}
\end{table}

Finally we summarize $\bar{T}_C$ and $\bar{T}_H$ in Table \ref{tab:04} for all the models as a function of 
 some selected inclination angle. In this table, the intrinsic minimum temperature $T_{\rm min}$ and the maximum temperature $T_{\rm max}$ are given as a reference. Changing the inclination angle from the equator ($\theta = 0^{\circ}$) to the pole ($\theta = 90^{\circ}$, see from left to right in Table 4), $\bar{T}_H$ commonly increases, because the cold region in the equator is
 obscured for a polar observer (see Fig.\ref{fig:09}). 
 Though all the models share a similar temperature contrast between $\bar{T}_C$ and $\bar{T}_H$, the ratio of the area (Table 3) of the two spots holds the key to determine whether the single or two component fitting is more preferential.
 
\begin{table*}[hbt]
\caption{\label{tab:04}
Temperature of cold spot and hot spot, and average temperature given by the two component BB fittings as a function of the inclination angle. 
All units are in [keV].  In non-magnetized models (M000, m000),
 $\bar{T}_C$ and $\bar{T}_H$ is not given (--) simply because 
the surface temperature is uniform.
}
\begin{tabular}{l|cc|cc|cc|cc|cc}
Name & $T_{\rm min}$ & $T_{\rm max}$ & $\bar{T}_C(0^\circ)$  & $\bar{T}_H(0^\circ)$  & $\bar{T}_C(30^\circ)$  & $\bar{T}_H(30^\circ)$ & $\bar{T}_C(60^\circ)$  & $\bar{T}_H(60^\circ)$ & $\bar{T}_C(90^\circ)$  & $\bar{T}_H(90^\circ)$ \\
\hline   
   M000 &0.131&0.131&-&-&-&-&-&-&-&- \\
   m000 &0.131&0.131&-&-&-&-&-&-&-&- \\
\hline
   mauk &0.028&0.136&0.087&0.128&0.088&0.130&0.092&0.132&0.121&0.134\\
   maUk &0.017&0.159&0.095&0.150&0.097&0.152&0.101&0.154&0.129&0.155\\
   MaUk &0.019&0.188&0.077&0.176&0.077&0.177&0.158&0.180&0.171&0.181\\
   mAUk &0.016&0.161&0.094&0.155&0.095&0.155&0.098&0.156&0.128&0.157\\
\hline   
   maUS &0.022&0.157&0.088&0.144&0.090&0.147&0.095&0.149&0.111&0.151\\  
   mSUK &0.021&0.158&0.087&0.144&0.089&0.147&0.094&0.149&0.109&0.151\\
\end{tabular}
\end{table*}

 \section{Summary and Discussions}
Bearing in mind the application to high-B pulsars,
we have investigated 2D thermal evolutions of NSs. 
We paid particular attention to the influence 
 of different equilibrium configurations on the 
 surface temperature distributions. The equilibrium configurations were
 constructed in a systematic manner,
 in which both toroidal and poloidal magnetic fields are determined 
self-consistently with the inclusion of GR effects. 
To solve the 2D heat transfer inside the NS interior out to the crust, 
we have developed an implicit code based on a finite-difference scheme 
that deals with anisotropic
 thermal conductivity and relevant cooling processes in the 
 context of a standard cooling scenario. In agreement with previous 
 studies,
 the surface temperatures near the pole become higher than those 
 in the vicinity of the equator as a result of the heat-blanketing 
 effect. Our results showed that the ratio of the highest to
 the lowest surface temperatures changes maximally by one order of 
magnitude, depending on the equilibrium configurations. 
Despite such inhomogeneous temperature distributions, we found that the area of such hot and cold spots is so small that the simulated X-ray spectrum could be well reproduced by a single temperature BB fitting.
 We speculated that if a two-component BB fitting is needed to account for the observed spectrum,
 the toroidal magnetic field could be more widely distributed 
inside the NS interior than for models that only require 
 a single temperature BB fitting.

Comparing with the state-of-the-art 2D models 
\citep{geppert02,geppert04,perez06,pons07,aguilera08,
pons09,vigano13},  this study that is our first attempt 
 to join in the NS evolutionary calculations have a number of
 caveats to be improved. First of all, we only took into account
 the Ohmic dissipation for simplicity. On the timescale of 10$^4$ yr
 explored in this study, 
it would not have any significant effects on the thermal evolution, 
 however, the inclusion of the Hall effect and ambipolar 
 diffusion (e.g., \citet{aguilera08,vigano13}) is inevitable 
 for studying the subsequent evolution to compare with observations.
  If direct Urca processes
 mediated by hyperons were taken into account \citep{lattimer91,
prakash92}, the neutrino luminosity could be significantly 
enhanced (by 5-6 orders of 
magnitudes) compared to that of the standard cooling scenario.
To test the minimal cooling and enhanced cooling scenarios is 
 also a major undertaking. Superfluidity and superconductivity should 
 be included, which should 
 modify the heat capacity and neutrino 
emissivities \citep{kaminker01,andersson05,lander12}, and provide 
a new heat source in the crust \citep{tsuruta09}. Regarding the equilibrium configuration, 
  the electric current is assumed to vanish
 at the surface in the present scheme.
 The updated numerical scheme recently
 proposed by \citet{fujisawa12} can handle the non-vanishing 
 toroidal component there, which should affect the surface temperature
 distributions. We assumed that the magnetic axis is aligned with the rotational axis.
 To accurately deal with a misalignment that is thought to be 
a general feature of pulsars, we have to construct 3D equilibrium
configurations, the numerical method of which has not been established 
yet.

A comparison with observations is one of the most important 
 issues in the theoretical study of the thermal evolutions of NSs. 
Based on the state-of-the-art 2D models including both elaborate
 cooling rates and field-decay processes, \citet{perna13} have 
 recently investigated the X-ray specta and pulse profiles for a 
variety of initial magnetic field configurations. 
They pointed out that the simulated pulse profiles are sensitive to 
the field configurations 
(e.g., the dominance between the toroidal and poloidal fields). 
Owing to the alignment of the rotational axis and the magnetic 
 axis in our 2D models, such analysis is unfortunately 
beyond the scope of this work. In addition to the required 
 improvements for this work (e.g., simplified treatment of 
 field decay and cooling processes), more accurate prediction
 of the X-ray spectra is mandatory, in which effects of 
(energy-dependent) interstellar absorption, light deflection, and 
gravitational redshift are taken into account as in \citet{perna13}.
At the very least (before we will tackle on this subject in the future), 
 let us note that the intrinsic cooling curves of 
our models (namely without the observational corrections) are 
qualitatively consistent with the ones in \citet{perna13}.
The upper panel of Figure 1 in Perna et al. (2013) shows 
 the cooling curve from one of their representative models,
 in which a purely poloidal field ($B_0 = 10^{13}$ G) is assumed and 
similar microphysics (such as the cooling processes, thermal conductivity,
 heat capacity, and EOS) is taken as those in this work.
As shown, our models (the left panel of 
Figure 7 in this work) reproduce similar results, in which
 the hot and cold spots appear on the magnetic poles and equatorial 
regions, respectively.
  
Keeping our efforts to improve these important ingredients, 
our final goal is to construct a fully self-consistent simulation,
 in which the stellar configuration is determined in a self-consistent 
 manner under the influence of the magnetic field decay, heating and 
 cooling processes \footnote{Using data from recent core-collapse 
 supernova simulations that successfully produce neutrino-driven explosions (e.g., \citet{suwa10,bruenn13,bernhard12,takiwaki12}, see also
\citet{janka12,kotake12,kotake13} for recent review), we think it also important to study a proto-neutron star evolution in the multi-D context.}. This study, in which we developed 
 a new code including equilibrium configurations 
(albeit employing a very crude approximation of the microphysics and 
 field decay treatment) is nothing but a prelude, however,
 an important trial for us to take a very first step to
 the long and winding road.






\bigskip
We would like to thank  K. Kiuchi, S. Yamada, Y.Eriguchi and K. 
Makishima for fruitful discussions. NY is also grateful to D. Vigan\`{o}, J.A. Pons, and J.A. Miralles for their warm hospitality during his 
 stay in the University of Alicante and for their useful and insightful
comments on this work. This study was supported in part by the Grants in Aid 
for the Scientific Research from the Ministry of Education, Science and Culture of Japan (no. 2510-5510, 23540323, 23340069, and 24244036).

\appendix
\section{An implicit scheme for 2D 
 thermal diffusion calculations}

\begin{figure}[tbh]
\begin{center}
\includegraphics[width=80mm]{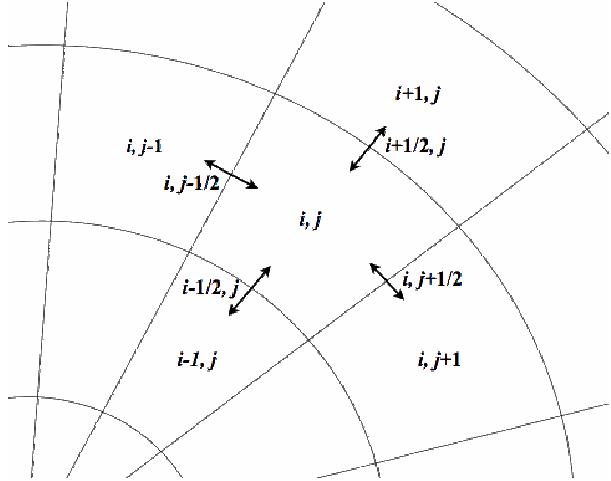}
\caption{Schematic picture of each numbering to the grids. Each vector shows the thermal flux for $F_{rr}$ or $F_{\theta \theta}$. The other fluxes  $F_{r \theta}$ and $F_{\theta r}$ are orthogonal to them, although we do not show in this figure.}
\label{fig:11}
\end{center}
\end{figure}

We briefly summarize an implicit scheme for 2D evolutionary calculations 
  developed in this work. Fig.~\ref{fig:11} illustrates spatial
 positions of a given point on the computational domain in our 2D code.
 The index $i$ and $j$ denotes the number of radial and lateral 
grid, respectively. Scalar valuables such as density ($\rho_{i,j}$),
 heat capacity (${c_{v~i, j}}$), redshift ($e^\phi_{i, j}$), 
temperature ($T_{i, j}$), and thermal conductivity ($\kappa_{i,j}$) are 
 defined at the center of the grid, while vector valuables
 such as the thermal flux are defined at the cell boundary 
($F_{i+1/2, j+1/2}$). The fluxes in the 2D computations consist of the 
 following four components,
\begin{eqnarray}
\label{eq:frr}
F_{rr} &=& -e^{-\phi} (\kappa_{rr} e^{-\Lambda} \partial_r \tilde T ),\\
F_{r \theta} 
&=& -e^{-\phi} (\frac{\kappa_{r\theta}}{r} e^{-\Lambda} \partial_\theta \tilde T),\\
\label{eq:fthth}
F_{\theta r} &=& -e^{-\phi} (\kappa_{\theta r} e^{-\Lambda} \partial_r \tilde T), \\
F_{\theta \theta} &=& -e^{-\phi} (\frac{\kappa_{\theta \theta}}{r} e^{-\Lambda} 
                                     \partial_\theta \tilde T).
\end{eqnarray}

Fixing hydrodynamic valuables such as density, pressure, and gravity as a 
 background, we solve the diffusion equation (Eq. \ref{eq:evo}) by an 
operator-splitting method. 

We solve the time evolution separately with the radial direction, the 
lateral direction, and the source term, respectively. The evolution 
 regarding the radial and lateral advection can be expressed, 
respectively as
\begin{eqnarray}
&\{ c_v e^\phi \}_{i, j}^{l+1}&\frac{T_{i, j}^{l+1} - T_{i, j}^l }{\Delta t}~dV_{i, j} 
\nonumber \\
&=& - \frac{ \{ e^{2\phi} F_{rr} \}_{i+1/2,j}^{l+1} \{ dS_r \}_{i+1/2, j} }{ dr_{i+1/2} }  \\ 
&&+ \frac{\{ e^{2\phi} F_{rr} \}_{i-1/2,j}^{l+1} \{ dS_r \}_{i-1/2, j} }{ dr_{i-1/2} }
\nonumber 
\end{eqnarray}
and 
\begin{eqnarray}
&\{ c_v e^\phi \}_{i, j}^{l+1} &\frac{T_{i, j}^{l+1} - T_{i, j}^l }{\Delta t}~dV_{i, j} 
\nonumber \\
&=& - \frac{\{ e^{2\phi} F_{\theta \theta} \}_{i,j+1/2}^{l+1} dS_{i,j+1/2} }{ r_{i} ~ d \theta} \\
&&  + \frac{\{ e^{2\phi} F_{\theta \theta} \}_{i,j-1/2}^{l+1} dS_{i,j-1/2} }{ r_{i} ~ d \theta },
\nonumber
\end{eqnarray}
where $dr_{i+1/2}= r_{i+1} - r_i$ is the radial grid interval, the index $l$ is the number of arbitral timestep, $d\theta$ is an equidistant angular 
 grid, $dV_{i, j}$ is a differential volume, and $\{ dS_r \}_{i+1/2, j} $
is the surface area between grid $(i, j)$ and $(i+1, j)$.

Finally we estimate the source term as,
\begin{eqnarray*}
&\{ c_v e^\phi \}_{i, j}^{l+1}& \frac{T_{i, j}^{l+1} - T_{i, j}^l }{\Delta t} ~dV_{i, j} 
\nonumber \\
&=& \{ e^{2\phi}_{i, j} (H_{i, j}-L_{i, j})  \}^{l+1}~dV_{i, j} \nonumber \\
&& - \frac{ \{ e^{2\phi} F_{r \theta} \}_{i+1/2,j}^{l} dS_{i+1/2,j} }{ dr_{i+1/2} } \nonumber \\ 
&&+ \frac{\{ e^{2\phi} F_{r \theta} \}_{i-1/2,j}^{l} dS_{i-1/2,j} }{ dr_{i-1/2} } \\
&& - \frac{\{ e^{2\phi} F_{\theta r} \}_{i,j+1/2}^{l} dS_{i,j+1/2} }{ r_{i} ~ d \theta} \nonumber \\
&& + \frac{\{ e^{2\phi} F_{\theta r} \}_{i,j-1/2}^{l} dS_{i,j-1/2} }{ r_{i} ~ d \theta }
\nonumber,
\end{eqnarray*}
where we do not solve the last four terms implicitly (such as $\partial_r \partial_\theta$ and $\partial_\theta \partial_r$)
that appear in the non-radial and non-lateral fluxes ($F_{r \theta}$ and $F_{\theta r}$), but treat them as a source term for simplicity. However,
  the first term is solved implicitly by iterations to get a 
numerical convergence.


\end{document}